\documentclass[aps,pra,twocolumn,superscriptaddress]{revtex4-1}
\usepackage{braket}
\usepackage{amsmath,amssymb,amsfonts}
\usepackage{amsthm}
\usepackage{graphicx}

\newtheorem{theorem}{Theorem}[section]
\newtheorem{lemma}[theorem]{Lemma}

\newcommand{\average}[1]{\mbox{$\langle #1 \rangle$}}

\newcommand{\norm}[1]{\left|\left|#1\right|\right|}

\newcommand{\be}{\begin{equation}}
\newcommand{\ee}{\end{equation}}
\newcommand{\ba}{\begin{eqnarray}}
\newcommand{\ea}{\end{eqnarray}}
\newcommand{\ban}{\begin{eqnarray*}}
\newcommand{\ean}{\end{eqnarray*}}
\newcommand{\one}{\leavevmode\hbox{\small1\normalsize\kern-.33em1}}

\usepackage{color}

\begin{document}

\title{Device-independent parallel self-testing of two singlets}
\author{Xingyao Wu}
\affiliation{Centre for Quantum Technologies, National University of Singapore, 3 Science Drive 2, 117543, Singapore}
\author{Jean-Daniel Bancal}
\affiliation{Centre for Quantum Technologies, National University of Singapore, 3 Science Drive 2, 117543, Singapore}
\author{Matthew McKague}
\affiliation{Dodd-Walls Centre for Photonic and Quantum Technologies, Department of Computer Science, University of Otago, Dunedin 9016, New Zealand}
\author{Valerio Scarani}
\affiliation{Centre for Quantum Technologies, National University of Singapore, 3 Science Drive 2, 117543, Singapore}
\affiliation{Department of Physics, National University of Singapore, 2 Science Drive 3, 117542, Singapore}

\begin{abstract}
Device-independent self-testing is the possibility of certifying the quantum state and the measurements, up to local isometries, using only the statistics observed by querying uncharacterized local devices. In this paper, we study parallel self-testing of two maximally entangled pairs of qubits: in particular, the local tensor product structure is not assumed but derived. We prove two criteria that achieve the desired result: a double use of the Clauser-Horne-Shimony-Holt inequality and the $3\times 3$ Magic Square game. This demonstrate that the magic square game can only be perfectly won by measureing a two-singlets state. The tolerance to noise is well within reach of state-of-the-art experiments.
\end{abstract}

\begin{widetext}
\maketitle
\end{widetext}

\textit{Introduction. --} State tomography is an almost necessary property of the notion of ``state": it must be possible, and indeed it is in quantum theory, to reconstruct the state fully by performing a suitable set of well-chosen measurements. In 1992, Popescu and Rohrlich proved a surprising result \cite{pr92}: the observation of the maximal quantum violation of the Clauser-Horne-Shimony-Holt inequality~\cite{CHSH}, $CHSH=2\sqrt{2}$ in the usual writing, certifies that the state must be a two-qubit maximally entangled state (``singlet") and that the two measurements on each side must be complementary (see also \cite{sw87,Tsirelson93}). In other words, even in a black box scenario in which nothing is known \textit{a priori} about the degrees of freedom under study, if some suitable statistics are observed the state and the measurements are characterized uniquely (up to local isometries, that is, up to local unitaries and unmeasured degrees of freedom). This fact is nowadays known as \textit{device-independent self-testing}. The name `self-testing' was proposed by Mayers and Yao when they demonstrated a similar criterion~\cite{MayersYao2004}; `device-independent' was coined in \cite{acin2007} and refers to certifications that do not rely on any knowledge of the Hilbert space dimension nor of the measurements, but only on the correctness of quantum theory and the evidence of a loophole-free Bell violation (see \cite{slovaca} and references therein).

The study of self-testing was boosted in the past few years. The original proofs were greatly simplified~\cite{McKagueThesis,McKague2012Robust} and several more states were found to be self-testable: all multipartite graph states~\cite{matthew}, all pure non-maximally entangled two-qubit states~\cite{Yang2013Robust,Bamps2014}, some other three-partite states that are not graph states~\cite{wstate1,wstate2}, one relevant two-qutrit state~\cite{swapshort}, optimal states for XOR games~\cite{millershi}... Besides, by building on self-testing, it was recently proved that interactive proofs for quantum computation \cite{RUV,mckague13} and arbitrary randomness amplification \cite{coudron,shi} are in principle possible with quantum resources.

All the works in this last series provide \textit{robustness} bounds: they can quantify how close the actual state is to the desired one, or how the device's performances are affected, when the observed statistics differ from the ideal ones, e.g. $CHSH=2\sqrt{2}(1-\epsilon)$. However, there are important areas for improvement. The techniques that achieve very good robustness $\epsilon\gtrsim 10\%$~\cite{swapshort,swaplong} apply under the assumption that subsequent uses of the black-boxes are independent and identically distributed (i.i.d.) processes. The works on interactive proofs and randomness amplification developed tools to avoid the i.i.d. assumption, but at the price of a dramatic drop in robustness. For instance, the pioneering proof of Reichardt, Unger and Vazirani \cite{RUV} is built on a i.i.d. self-testing scheme with robustness $\epsilon\sim 10^{-5}$ \footnote{In Eq.~(41) of \cite{swaplong} we find the estimate $||\psi-\bar{\psi}\otimes\mathrm{junk}||^2\leq 72284\times 4\epsilon$ (notice that $\epsilon$ in that paper is equal to $2\sqrt{2}\epsilon$ in our notation). Even leaving self-testing aside, the algebraic maximum of the l.h.s. is 2; therefore the estimate is trivial for $\epsilon\geq\frac{1}{144568}$.}, and the robustness of the overall protocol decreases as a very high power polynomial in the number of runs.

In this work, we contribute to filling the gap, coming from the side of the high robustness proofs and dropping the i.i.d. assumption. Specifically, we are going to provide two criteria using bipartite four output correlations that self-test the presence of \textit{two} singlets in a black box. The first criterion is \textit{Double CHSH}: one takes four inputs and four outputs on each side, interpret them as two pairs of bits, and run two CHSH tests. Interestingly, the mere fact of observing $CHSH=2\sqrt{2}$ in both tests is not sufficient to self-test two singlets: one has to have a closer look to the observed correlations (see Appendix \ref{proof2chsh}). The second criterion uses the $3\times 3$ \textit{Magic Square game} \cite{mermin,cabello}, which can be won perfectly with two singlets. Its self-testing character was not known prior to this paper, even in the perfect case \footnote{In a former draft of this paper, we erroneously stated that the proof for the ideal case follows from: L. Mancinska, LNCS \textbf{8808}, 200 (2014); arXiv:1506.07080. In reality, this paper deals only with \textit{weakly projective} pseudo-telepathy games: for each question that Bob is asked, there is at least one question for Alice so that, to win, Bob's answer is a function of Alice's answer.  (or the same thing, with Alice and Bob reversed). For the magic square game, this is not the case, since Bob's answer has two parts, and for any pair of questions for Alice and Bob, there is at least one bit of freedom in Bob's answer.}. For both criteria, we are going to provide robustness bounds that can be used to assess state-of-the-art experiments. We stress that the boxes see only the four-valued input and produce a four-valued output: we are working in the \textit{two-copy parallel-repetition} scenario. This is more general than the sequential scenario studied previously, in which the boxes are fed with one bit and produce one bit each time, and in which the future may depend on the past but not the other way round \cite{RUV}.

We start by setting up the scenario and summarizing the task of self-testing.

\begin{figure}
\includegraphics[width=0.4\textwidth]{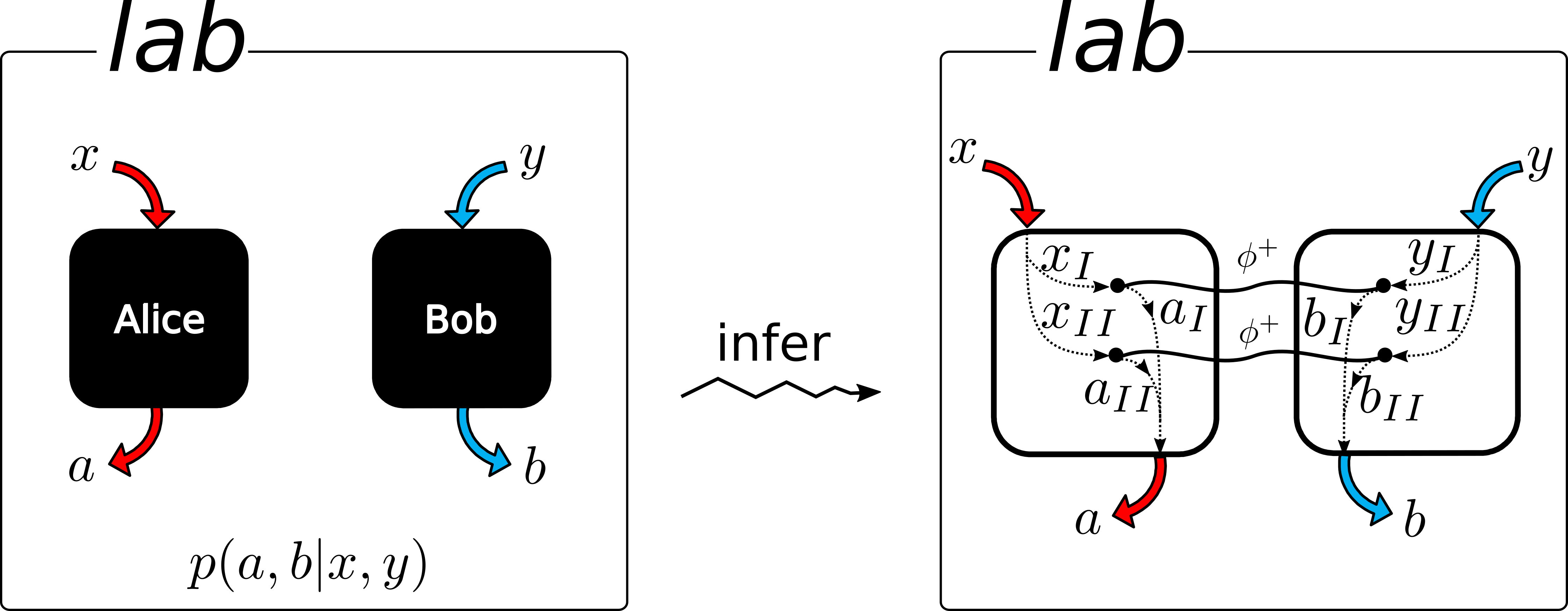}
\caption{Joint certification of two maximally-entangled qubit states. Two parties have a box with three or four possible settings and four possible outcomes. Can they demonstrate that they share two singlet states? In the case of four inputs, can they interpret their input and output quads as pairs of bits, which individually address each singlet?}\label{fig:setup}
\end{figure}

\textit{Measurement scenario. --} We consider a physical system in the state $\rho_{\mathcal{A}\mathcal{B}}$ shared between two separated parties $\mathcal{A}$ and $\mathcal{B}$. The possible measurements on sub-system $\mathcal{A}$ ($\mathcal{B}$) are labelled by $x$ ($y$) and their outcome by $a$ ($b$). Since the Hilbert space dimension is not bounded, we can assume that the measurements on the state are projective. This means that for any setting of Alice there exist an otherwise unknown set of projectors $\{\Pi_{a|x}\}$, and the same for Bob with $\{\Pi_{b|y}\}$. After sufficient runs of the process, one can estimate
\begin{align*}
P(a,b|x,y) = \average{\Pi_{a|x}\Pi_{b|y}} = \text{Tr} \big[\rho_{\mathcal{A}\mathcal{B}}\Pi_{a|x} \Pi_{b|y}\big].
\end{align*}
These \textit{observed correlations}, together with the fact that they must be obtainable from local measurements on a quantum state, are the constraints in the derivation of self-testing bounds.

\textit{Robust self-testing. --} For a bipartite system, we say that an unknown quantum state $\rho$ is self-tested into a well-defined quantum state $\ket{\psi}$ if there exist a local isometry $\Phi=\Phi_{\mathcal{AA'}}\otimes\Phi_{\mathcal{BB'}}$ such that
\begin{align}\label{eq:definitionSelftesting}
\Phi\left(\rho_{\mathcal{A}\mathcal{B}}\otimes \xi_{\mathcal{A}^{'}\mathcal{B}^{'}}\right)\Phi^{\dagger}&=\rho_{\mathcal{AB}}^{junk}\otimes \ket{\psi}_{\mathcal{A}^{'}\mathcal{B}^{'}}\bra{\psi},
\end{align} where $\mathcal{A}'$ and $\mathcal{B}'$ are ancillas of the same dimension as the subsystems of $\ket{\psi}$, prepared in the state $\xi$. We'll need to find the suitable $\Phi$.

Only some $P(a,b|x,y)$ may self-test a state exactly; otherwise, a suitable $\Phi$ could still leave the auxiliary system in a state relatively close to the target one. We measure this closeness with the fidelity
\begin{align}
F=\bra{\psi}\rho_{\mathcal{A}'\mathcal{B}'}\ket{\psi},
\end{align} where $\rho_{\mathcal{A}'\mathcal{B}'}=\text{Tr}_{\mathcal{A}\mathcal{B}}\big[ {\Phi\,\rho_{\mathcal{A}\mathcal{B}}\otimes\xi_{\mathcal{A}^{'}\mathcal{B}^{'}}\,\Phi^\dagger}\big]$ is the partial state of the ancillas after the isometry.

In order to obtain a robustness bound, i.e., a lower bound on $F$, we follow the method of Refs \cite{swapshort,swaplong}. In a nutshell, it relies on expressing $\Phi$ as a linear combination of the measurement operators, which must be proved to be unitary. Then $F$ becomes the average of a linear combination of operator monomials $\mathcal{C}=(\Pi_{a|x}, \Pi_{b|y}, \Pi_{a|x}\Pi_{b|y},\Pi_{a|x}\Pi_{a'|x'}\Pi_{b|y}, ...)$, to be evaluated on the observed correlations $P(a,b|x,y)$. The latter however don't determine terms like $\langle\Pi_{a|x}\Pi_{a'|x'}\rangle$. Thus we are led to solve the following SDP:
\begin{equation}\label{sdp}
\begin{array}{lll}
\min & F(\mathcal{C}) \\\\
\text{s.t.} & \Gamma\geq 0,\\\\
&P(a,b|x,y)=\mathrm{Tr}_{AB}[\rho_{AB}\Pi_{a|x}\Pi_{b|y}], \\
\end{array}
\end{equation} 
where $\Gamma=\average{\mathcal{C^\dagger}\mathcal{C}}$ is a matrix of moments. It was proved by Navacsu\'es, Pironio and Ac\'{\i}n (NPA) that, for any choice of monomials $\mathcal{C}$, the constraint $\Gamma\geq 0$ is a relaxation of the requirement that the correlations must be of quantum origin \cite{npa}. In other words, the SDP provides a lower bound on the certifiable fidelity.

This method is therefore algorithmic once the isometry is chosen. In order to tackle the two cases of interest, \textit{Double CHSH} and \textit{Magic Square}, we need to construct isometries that give good bounds (we won't prove them to be optimal).

\textit{Double CHSH. --} For this criterion we need four inputs and four outputs on each party, i.e. $x,y,a,b\in\{0,1,2,3\}$. The idea is to formally split these inputs and outputs as if they belonged to two separate CHSH tests, each denoted by $j\in\{I,II\}$ (see Figure~\ref{fig:setup}): 
\ba
x=2x_I+x_{II}&,& y=2y_I+y_{II}\,,\\
a=2a_I+a_{II}&,& b=2b_I+b_{II}
\ea
with $x_j,y_j,a_j,b_j\in\{0,1\}$.


In order to construct the isometry, we can assume that the ancilla systems are two qubits on each side $\mathcal{A}'=\{{\mathcal{A}}'_{I},{\mathcal{A}}'_{II}\}$, $\mathcal{B}'=\{{\mathcal{B}}'_{I},{\mathcal{B}}'_{II}\}$. For the construction of the swap operator, as in previous works~\cite{swapshort,swaplong} we take inspiration from the ideal case. In the ideal case, the systems are two qubits on each side, $\mathcal{A}=\{{\mathcal{A}}_{I},{\mathcal{A}}_{II}\}$, $\mathcal{B}=\{{\mathcal{B}}_{I},{\mathcal{B}}_{II}\}$. If we write the ideal state of the system as a rotated singlet
\begin{align}
\label{2chshstate}
\ket{\psi}=\big(\cos(\frac{\pi}{8})\ket{\Phi^+}+\sin(\frac{\pi}{8})\ket{\Psi^-}\big)^{\otimes2},
\end{align} the optimal CHSH measurements are the same on both parties, namely
\ba \textrm{$\sigma_z$ for $x_j,y_j=0$, and $\sigma_x$ for $x_j,y_j=1$}\,.\label{chshsettings}\ea
The swapping between one qubit $\mathcal{A}_j$ and its ancilla $\mathcal{A'}_j$ is then $\mathcal{S}_j=UVU$ with
\ban
U=\one\otimes\pi_{z=0}+\sigma_{x}\otimes\pi_{z=1}\,\\
V=\pi_{z=0}\otimes\one+\pi_{z=1}\otimes\sigma_{x}
\ean
where the $\pi_{z=0/1}$ are the projectors on the eigenstates of $\sigma_z$. The ideal swap for the whole setup is $\mathcal{S}_{I}\otimes\mathcal{S}_{II}$ on Alice's side and the same construction on Bob's. From now onwards, we omit the tensor product sign to make the notation less cumbersome. If we prepare the ancillas in the state $\ket{00}_{{\mathcal{A'}}}$, we find explicitly
\ba
\mathcal{S}_{I}\mathcal{S}_{II}\ket{00}_{{\mathcal{A'}}}&=& \ket{00}_{{\mathcal{A'}}} \pi^{\mathcal{A}_I}_{z=0}\pi^{\mathcal{A}_{II}}_{z=0}\nonumber\\
&+& \ket{01}_{{\mathcal{A'}}} \sigma_x^{\mathcal{A}_{II}}\,\pi^{\mathcal{A}_I}_{z=0}\pi^{\mathcal{A}_{II}}_{z=1}\nonumber\\
&+& \ket{10}_{{\mathcal{A'}}} \sigma_x^{\mathcal{A}_I}\,\pi^{\mathcal{A}_I}_{z=1}\pi^{\mathcal{A}_{II}}_{z=0}\nonumber\\
&+& \ket{11}_{{\mathcal{A'}}} \sigma_x^{\mathcal{A}_I}\sigma_x^{\mathcal{A}_{II}}\,\pi^{\mathcal{A}_I}_{z=1}\pi^{\mathcal{A}_{II}}_{z=1}.\label{SS1}
\ea
Note that we used here the fact that in the perfect case, $I$ and $II$ are well-defined subsystems, hence their operators commute. Finally we replace the operators on ${\mathcal{A}_I},{\mathcal{A}_{II}}$ with the expressions involving the $\{\Pi_{a|x}\}$. Again the ideal case suggests the replacement:
\ban
\begin{array}{lcl}
\Pi_{a|x=0}\stackrel{id}{=}\pi_{z=a_I}^{\mathcal{A}_I}\,\pi_{z=a_{II}}^{\mathcal{A}_{II}}&,&
\Pi_{a|x=1}\stackrel{id}{=}\pi_{x=a_I}^{\mathcal{A}_I}\,\pi_{z=a_{II}}^{\mathcal{A}_{II}}\,,\\
\Pi_{a|x=2}\stackrel{id}{=}\pi_{z=a_I}^{\mathcal{A}_I}\,\pi_{x=a_{II}}^{\mathcal{A}_{II}}&,&
\Pi_{a|x=3}\stackrel{id}{=}\pi_{x=a_I}^{\mathcal{A}_I}\,\pi_{x=a_{II}}^{\mathcal{A}_{II}}\,,
\end{array}
\ean
whence
\ba
\begin{array}{lcl}
	\label{def}
\pi^{\mathcal{A}_I}_{z=s}\pi^{\mathcal{A}_{II}}_{z=t}&\longrightarrow& \Pi_{2s+t|0}\,,\\
\sigma_x^{\mathcal{A}_I} &\longrightarrow& \Pi_{0|3}+\Pi_{1|3}-\Pi_{2|3}-\Pi_{3|3}\,,\\
\sigma_x^{\mathcal{A}_{II}}&\longrightarrow& \Pi_{0|3}-\Pi_{1|3}+\Pi_{2|3}-\Pi_{3|3}\,,\\
\sigma_x^{\mathcal{A}_I}\,\sigma_x^{\mathcal{A}_{II}}&\longrightarrow& \Pi_{0|3}-\Pi_{1|3}-\Pi_{2|3}+\Pi_{3|3}\,.
\end{array}
\ea
Notice that we could have used $\{\Pi_{a|x=1}\}$ to replace $\sigma_x^{\mathcal{A}_I}$ and $\{\Pi_{a|x=2}\}$ to replace $\sigma_x^{\mathcal{A}_{II}}$; but one choice is sufficient, and the fact that these operators don't appear in the swap operator is not a problem (they will still appear in semi-definite matrix $\Gamma$).

In summary, the following operator acts as a swap on Alice's side whenever the expected measurements are performed:
\ba
\mathcal{S}_{\mathcal{A}\mathcal{A'}}\ket{00}_{\mathcal{A'}}&=&\ket{00}_{\mathcal{A'}} \Pi_{0|0} \label{SS2}\\
&+& \ket{01}_{\mathcal{A'}} (\Pi_{0|3}-\Pi_{1|3}+\Pi_{2|3}-\Pi_{3|3})\,\Pi_{1|0}\nonumber\\
&+& \ket{10}_{\mathcal{A'}} (\Pi_{0|3}+\Pi_{1|3}-\Pi_{2|3}-\Pi_{3|3})\,\Pi_{2|0}\nonumber\\
&+& \ket{11}_{\mathcal{A'}} (\Pi_{0|3}-\Pi_{1|3}-\Pi_{2|3}+\Pi_{3|3})\,\Pi_{3|0}\,.\nonumber
\ea
We thus choose to use this relation as the definition of the swap operator in the non-ideal case as well. In order to check whether this is possible, we have to check that this definition is compatible with a unitary operator $\mathcal{S}_{\mathcal{A}\mathcal{A'}}$. Having fixed the initial state of the ancillas, this can be done by checking that the map $\ket{\psi}\longrightarrow
\mathcal{S}_{\mathcal{A}\mathcal{A'}}\ket{\psi}_{\mathcal{A}}\otimes\ket{00}_{\mathcal{A'}}$ preserves the scalar product, which is readily confirmed (see also Appendix \ref{swapunitary}).

Because of the choice of state \eqref{2chshstate}, we can use exactly the same construction on Bob's side, with $\Pi_{a|x}$ replaced by the operators $\{\Pi_{b|y}\}$. Having defined the isometry, we can now proceed to the optimisation~\eqref{sdp} for the \textit{Double CHSH} correlations. The results will be presented at the end of the paper, after describing the construction of the isometry for the \textit{Magic Square} correlations.

\textit{Magic Square. --} Though not as popular as CHSH, the magic square game is well known in the literature and we refer to it for its introduction \cite{mermin, cabello, broadbent}. For our purpose here, it is sufficient to describe the ideal implementation. Alice and Bob share two qubits each, and can each perform one of three measurements ($x,y\in\{0,1,2\}$). In a basis that is convenient to construct the swap, and using the notation $c=a,b$ for Alice and Bob respectively, the possible measurements are
\ba
\begin{array}{l}
\Pi_{c|0}\,=\,\pi_{z=c_{I}}\otimes \pi_{z=c_{II}}\,,\\
\Pi_{c|1}\,=\,\pi_{x=c_{I}}\otimes \pi_{x=c_{II}}\,,\\
\Pi_{c=0,1|2} = \ket{\chi_{\pm}}\bra{\chi_{\pm}}\,,\, \Pi_{c=2,3|2} = \ket{\chi'_{\pm}}\bra{\chi'_{\pm}}
\end{array}\label{magicsettings}\ea
with $\ket{\chi_{\pm}}=\frac{1}{\sqrt{2}}(\ket{\Phi^-}\pm\ket{\Psi^+})$ and $\ket{\chi'_{\pm}}=\frac{1}{\sqrt{2}}(\ket{\Phi^+}\pm\ket{\Psi^-})$ maximally entangled states. The statistics of these measurements on the state
\begin{align}
\label{msbstate}
\ket{\psi}=\ket{\Phi^+}_{{\mathcal{A}_I} {\mathcal{B}_I}}\otimes \frac{\ket{\Phi^+}+\ket{\Psi^-}}{\sqrt{2}}\,_{{\mathcal{A}_{II}} {\mathcal{B}_{II}}}
\end{align} cannot be reproduced with shared randomness and win the magic square game with certainty.

We are going to prove that these statistics actually self-test two singlets. The ideal case can be done analytically. With the same techniques as \cite{MM15} one obtains the following robustness bound: there exists an isometry $\Phi$ and junk state $\ket{junk}$ such that
\begin{equation}\label{analyticmagic}
\norm{
    \Phi(\ket{\psi^{\prime}})
    -
    \ket{junk}
    \ket{\psi}
} 
\leq
10.13
(2\epsilon)^{\frac{1}{4}}
\end{equation}
where $\epsilon$ is the maximum error in the expectation value for any measurement as compared to the perfect case. Since the l.h.s. is at most $\sqrt{2}$, the bound becomes trivial for $\epsilon\gtrsim 2\times 10^{-4}$. So this bound, which can be handy for theoretical work, can tolerates a smaller error than the SDP optimisation whose result is discussed in the following section. All the details, including robustness bounds on the measurements, are found in Appendix \ref{appmm}.

In the main text, we focus on achieving robust self-testing with the SDP optimisation, we shall need to provide the operator $\mathcal{S}_{\mathcal{A}\mathcal{A'}}$ that ideally swaps the system and the ancilla. In fact, we can borrow the previous one \eqref{SS2}, with the replacement $\Pi_{a|3}\longrightarrow\Pi_{a|1}$: indeed, the construction uses only the operators that in the ideal case read $\pi_{z=a_{I}}\otimes \pi_{z=a_{II}}$ and $\pi_{x=a_{I}}\otimes \pi_{x=a_{II}}$, and with our notations we have them also here, for both Alice and Bob.




\textit{Robustness bounds from SDP optimisation. --} Finally we discuss the SDP optimisation. For the \textit{Double CHSH}, we consider the correlations $P(a,b|x,y)$ that are obtained if the ideal measurements \eqref{chshsettings} are performed on the non-ideal state
\begin{align}\label{statechsh}
\rho(\varepsilon)=\bigg((1-\varepsilon)\ket{\psi_0}\bra{\psi_0}+\varepsilon\one/4\bigg)^{\otimes 2},
\end{align}
where $\ket{\psi_0}=\cos(\frac{\pi}{8})\ket{\Phi^+}+\sin(\frac{\pi}{8})\ket{\Psi^-}$. For this state, one would see $CHSH_{I}=CHSH_{II}=2\sqrt{2}(1-\varepsilon)$.
Similarly, for the \textit{Magic Square}, we take the correlations that are obtained if the ideal measurements \eqref{magicsettings} are performed on the non-ideal state \begin{align}\label{statemagic}
\rho(\varepsilon)=&\big((1-\varepsilon)\ket{\Phi^+}\bra{\Phi^+}+\varepsilon\one/4\big)\nonumber\\
&\otimes\big((1-\varepsilon)\ket{\chi'_+}\bra{\chi'_+}+\varepsilon\one/4\big).
\end{align} As compared to the definition of $\epsilon$ used in Eq.~\eqref{analyticmagic}, for this state it holds $1-\epsilon=(1-\varepsilon)^2$ i.e. $\varepsilon\approx\epsilon/2$.

\begin{figure}[htbp!]
	\centering
    \includegraphics[width=0.48\textwidth]{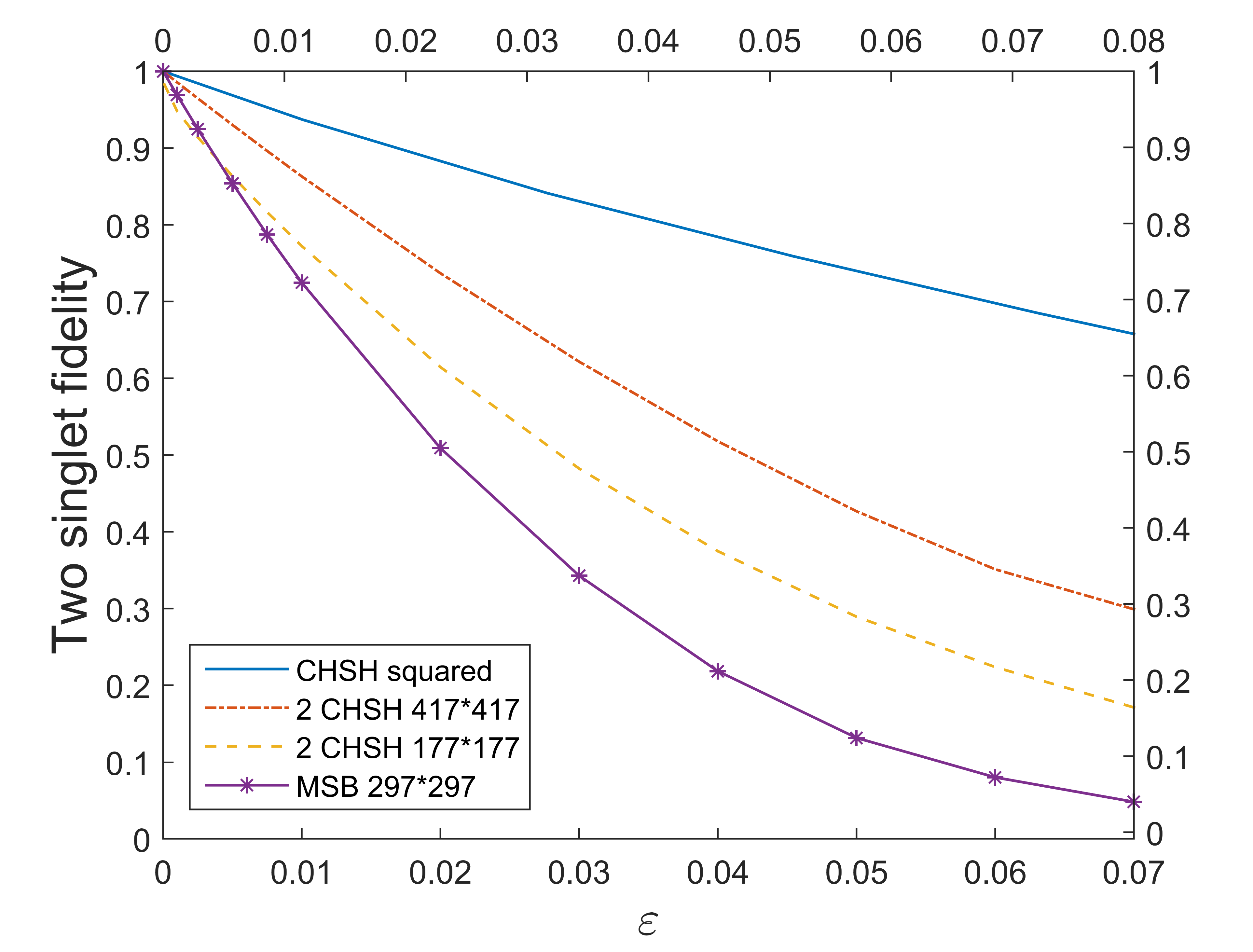}
	\caption{(Color online) Bound on the fidelity of two singlet states as a function of the parameter $\varepsilon$ for Double CHSH [assuming the statistics of the state \eqref{statechsh}] and for Magic Square [assuming the statistics of the state \eqref{statemagic}]. The solid line represents the squared value of fidelity from single-copy self-testing. The sizes of the SDP matrices are noted in the legend.}
	\label{fig:swapcurve}
\end{figure}

The results of the optimisation for different values of $\varepsilon$ is shown in Figure~\ref{fig:swapcurve}. For both tests we have used the set of moments
\ba
\mathcal{C}&=&\Big\{1,\{\Pi_{a|x}\},\{\Pi_{b|y}\},
\{\Pi_{a|x}\Pi_{b|y}\},\nonumber\\&& \{\Pi_{a|0}\Pi_{a'|3}\Pi_{b|0}\},\{\Pi_{a|0}\Pi_{b|0}\Pi_{b'|3}\}\Big\}
\ea which consists of $1+16+16+256+64+64=417$ elements for Double CHSH and of $1+12+12+144+64+64=297$ elements for Magic Square. Two other curves are plotted for comparison. One is the Double CHSH certification with a smaller $\mathcal{C}$, in which the 256 $\Pi_{a|x}\Pi_{b|y}$ are replaced by the 16 $\Pi_{a|0}\Pi_{b|0}$. This is to show that a smaller matrix gives already a decent bound, albeit obviously less tight. The other curve plots $F_1^2$, where $F_1$ is the bound on the one-singlet fidelity for the state $(1-\varepsilon)\ket{\psi_0}+\varepsilon\one/4$ \cite{swaplong}.

The Magic Square criterion appears to be less robust than Double CHSH, even comparing with the bound that uses a smaller matrix (for most $\varepsilon$) \footnote{Notice that the comparison in Fig.~\ref{fig:swapcurve} is based on assuming similarly imperfect resources, i.e. states \eqref{statechsh} and \eqref{statemagic}. If we were to plot as a function of $\epsilon$, the deviation in observed values, the curve for Magic Square would be almost identical to that of Double CHSH for 417 elements.}. Also, the curves suggest that the parallel scenario is different from two independent copies. We stress however that definite and general comparisons cannot be drawn, insofar as the bounds are derived for the $P(a,b|x,y)$ defined above and are not guaranteed to be tight. Even so, we prove that a relevant self-testing of two singlets can be certified with robustness up to a few percent. Let us therefore finish by applying our bounds to an extrapolation of the performance of recent experiments.

Three ``loophole-free'' Bell tests have been reported recently. While the photonic ones \cite{zeilinger,kwiat} used non-maximally entangled states, making a comparison with our present work meaningless, the Delft experiment \cite{delft} did target the singlet as the ideal state and reported a violation $CHSH\approx 2.42$. Such a value would correspond to $\varepsilon\approx 0.14$. Then our SDP give $F_1\geq 0.64$ and $F\geq 0.09$ (with the $417\times 417$ matrix, outside the axes of Fig.~\ref{fig:swapcurve}). This value of $F$ is not yet of practical interest. However, if we move away from loophole-free implementations, the best violation of CHSH reported after post-selection is $CHSH=2.8276$ \cite{poh}. This would correspond to $\varepsilon\approx 3\times 10^{-4}$, for which our SDP certifies $F_1\geq 0.999$ and $F\geq 0.996$. These are just quick estimates, insofar as we have neglected finite-sample effects; but they do show that, thanks to our method, state-of-the-art experiments already allow for a good self-testing of the singlet in the single-copy case, as well as in the two copies case \footnote{It is customary to benchmark singlet fidelity to that of separable states, which would be $F_1=0.5$ and $F=0.25$; however, since our estimates are obtained by violating a Bell inequality, the state is surely not separable.}.

\textit{Conclusion. --} We have presented two criteria that self-test two singlets in parallel in a device-independent way. Besides proving self-testing in the ideal case, we provided robustness bounds that give non-trivial estimates for existing experiments.

\textit{Acknowledgements. --} We thank Michal Hajdu\v{s}ek for suggesting the idea of the magic square game; Laura Man\v{c}inska, Thomas Vidick and Cong Wan for useful comments on an earlier draft.
This work is funded by the Singapore Ministry of Education (partly through the Academic Research Fund Tier 3 MOE2012-T3-1-009), the National Research Foundation of Singapore, and by the Dodd-Walls Centre for Photonic and Quantum Technologies.

\newpage

\appendix

\section{Double CHSH self-tests two singlets in the ideal case}
\label{proof2chsh}

That the mere observation of two CHSH violations may not self-test two singlets is obvious from the following counterexample. Suppose that only the settings $x=0 = (x_{I}=0,x_{II}=0)$ and $x=3=(x_{I}=1,x_{II}=1)$ are used, and similarly for Bob: these settings are sufficient to evaluate two CHSH inequalities, and both inequalities can be maximimally violated by just measuring one singlet and outputting the result twice.  From these statistics we know that we can create two isometries that each extract a singlet, but they would be the same isometry!  Clearly we need some form of independence between the two subtests to guarantee that the extracted singlets are different.  Here we will break down the first CHSH subtest into two tests, depending on $\mathcal{A}$'s question on the second subtest, and similarly for the second subtest.  This will enforce independence between the two subtests.

First, some definitions.  For $q \in (0,1)^n$ a bit string with $q_j$ the $j$th bit of the string, and $\{X^{\prime}_j\}_{j=1}^n$ some operators, define
\begin{equation}
X^{\prime q} :=
\prod_j
X^{\prime q_j}_j
\end{equation}
and analogously for $Z^{\prime p}$.  Next, for an operator $X$ we define $|X| :=  \sqrt{X^2}$, which is an operator with the same eigenvectors as $X$, but with eigenvalues equal to the absolute value of each eigenvalue of $X$. 

We will make use of a lemma from \cite{MM15}, stated below for our special case:
\begin{lemma}[\cite{MM15}, Lemma 4]
\label{lemma:graphstateselftestconditions}
Let $R(u)$ be the function that takes a 4-bit sequence $s_1 s_2 s_3 s_4$ to $s_3 s_4 s_1 s_2$, and suppose that $\ket{\psi^{\prime}} \in \mathcal{H}_\mathcal{A} \otimes \mathcal{H}_\mathcal{B}$ is a normalized state, and $\{X^{\prime}_1, X^{\prime}_2, Z^{\prime}_1, Z^{\prime}_2\}$ and  $\{X^{\prime}_3, X^{\prime}_4, Z^{\prime}_3, Z^{\prime}_4\}$ are unitary, Hermitian operators on $\mathcal{H}_\mathcal{A}$ and $\mathcal{H}_\mathcal{B}$, respectively, such that for any $s,t \in (0,1)^{4}$
\begin{equation}
\label{eq:XZanticommute9}
		X^{\prime s} Z^{\prime t} \ket{\psi^{\prime}}
		=
		(-1)^{s \cdot t}
		Z^{\prime t} X^{\prime s} \ket{\psi^{\prime}}
\end{equation}
and
\begin{equation}
\label{eq:XZswapgeneral9}
	X^{\prime s} \ket{\psi^{\prime}}
	=
	(-1)^{P(s)}
	Z^{\prime R(s)} \ket{\psi^{\prime}}
\end{equation}
Then there exists an isometry $\Phi = \Phi_{\mathcal{A}} \otimes \Phi_{\mathcal{B}}$ and a state $\ket{junk} \in \mathcal{H}_\mathcal{A} \otimes \mathcal{H}_\mathcal{B}$ such that for any $p,q \in (0,1)^{n}$
\begin{eqnarray}
\nonumber
    \Phi(
        X^{\prime q}
        Z^{\prime p}
        \ket{\psi^{\prime}}
    )
    =
    \ket{junk}
    X^q
    Z^p
    \ket{\psi}    
\end{eqnarray}
where $\ket{\psi}$ is the $4$-qubit state
\begin{equation}
\ket{\psi} = 
\frac{1}{\sqrt{2}}
\left(
    \ket{0+} + \ket{1-}
\right)_{13}
\otimes
\frac{1}{\sqrt{2}}
\left(
    \ket{0+} + \ket{1-}
\right)_{24}.
\end{equation}
\end{lemma}

Now we fix some strategy of $\mathcal{A}$ and $\mathcal{B}$ for playing two copies of CHSH.  They hold onto some joint state $\ket{\psi^{\prime}}$ and perform projective measurements $\Pi^{\mathcal{A}}_{a| x}$ and $\Pi^{\mathcal{B}}_{b| z}$ for inputs $x$ and $z$ and outputs $a$ and $b$.  Next we define some operators for our convenience:
\begin{eqnarray}
Z^{\prime}_{1} & := & \Pi^{\mathcal{A}}_{0|0} + \Pi^{\mathcal{A}}_{1|0} - \Pi^{\mathcal{A}}_{2|0}- \Pi^{\mathcal{A}}_{3|0} \\
Z^{\prime}_{2} & := & \Pi^{\mathcal{A}}_{0|0} - \Pi^{\mathcal{A}}_{1|0} + \Pi^{\mathcal{A}}_{2|0} - \Pi^{\mathcal{A}}_{3|0} \\
X^{\prime}_1 & := & \Pi^{\mathcal{A}}_{0|3} + \Pi^{\mathcal{A}}_{1|3} - \Pi^{\mathcal{A}}_{2|3}- \Pi^{\mathcal{A}}_{3|3} \\
X^{\prime}_2 & := & \Pi^{\mathcal{A}}_{0|3} - \Pi^{\mathcal{A}}_{1|3} + \Pi^{\mathcal{A}}_{2|3} - \Pi^{\mathcal{A}}_{3|3} 
\end{eqnarray}
\begin{eqnarray}
Z^{\prime \prime}_{1} & := & \Pi^{\mathcal{A}}_{0|2} + \Pi^{\mathcal{A}}_{1|2} - \Pi^{\mathcal{A}}_{2|2}- \Pi^{\mathcal{A}}_{3|2} \\
Z^{\prime \prime}_{2} & := & \Pi^{\mathcal{A}}_{0|1} - \Pi^{\mathcal{A}}_{1|1} + \Pi^{\mathcal{A}}_{2|1} - \Pi^{\mathcal{A}}_{3|1} \\
X^{\prime \prime}_1 & := & \Pi^{\mathcal{A}}_{0|1} + \Pi^{\mathcal{A}}_{1|1} - \Pi^{\mathcal{A}}_{2|1}- \Pi^{\mathcal{A}}_{3|1} \\
X^{\prime \prime}_2 & := & \Pi^{\mathcal{A}}_{0|2} - \Pi^{\mathcal{A}}_{1|2} + \Pi^{\mathcal{A}}_{2|2} - \Pi^{\mathcal{A}}_{3|2} 
\end{eqnarray}
\begin{eqnarray}
\nonumber
2V^{\prime}_3 & := & \Pi^{\mathcal{B}}_{0|0} + \Pi^{\mathcal{B}}_{1|0} - \Pi^{\mathcal{B}}_{2|0}- \Pi^{\mathcal{B}}_{3|0} \\
& & + \Pi^{\mathcal{B}}_{0|2} + \Pi^{\mathcal{B}}_{1|2} - \Pi^{\mathcal{B}}_{2|2}- \Pi^{\mathcal{B}}_{3|2} \\
\nonumber
2V^{\prime}_4 & := & \Pi^{\mathcal{B}}_{0|0} - \Pi^{\mathcal{B}}_{1|0} + \Pi^{\mathcal{B}}_{2|0} - \Pi^{\mathcal{B}}_{3|0} \\
& & +\Pi^{\mathcal{B}}_{0|1} - \Pi^{\mathcal{B}}_{1|1} + \Pi^{\mathcal{B}}_{2|1} - \Pi^{\mathcal{B}}_{3|1} 
\end{eqnarray}
\begin{eqnarray}
\nonumber
2W^{\prime}_3 & := & \Pi^{\mathcal{B}}_{0|3} + \Pi^{\mathcal{B}}_{1|3} - \Pi^{\mathcal{B}}_{2|3}- \Pi^{\mathcal{B}}_{3|3} \\
& & + \Pi^{\mathcal{B}}_{0|1} + \Pi^{\mathcal{B}}_{1|1} - \Pi^{\mathcal{B}}_{2|1}- \Pi^{\mathcal{B}}_{3|1} \\
\nonumber
2W^{\prime}_4 & := & \Pi^{\mathcal{B}}_{0|3} - \Pi^{\mathcal{B}}_{1|3} + \Pi^{\mathcal{B}}_{2|3} - \Pi^{\mathcal{B}}_{3|3} \\
& &\Pi^{\mathcal{B}}_{0|2} - \Pi^{\mathcal{B}}_{1|2} + \Pi^{\mathcal{B}}_{2|2} - \Pi^{\mathcal{B}}_{3|2}
\end{eqnarray}

These can be understood as follows: $Z^{\prime}_1$ is the measurement operator that $\mathcal{A}$ applies to find the first output bit when queried with the question $0$ for both subtests.  $Z^{\prime}_2$ is the measurement operator for the second output bit in the same question.  The other operators for $\mathcal{A}$'s actions are similarly defined.  For $\mathcal{B}$, however, we set $V^{\prime}_1$ to be the measurement operator for the first output bit when question $0$ is asked for the first subtest, forgetting what the second question is but assuming that it is uniformly distributed.  The other operators for $\mathcal{B}$ are similarly defined.

In all cases sub-indices 1 and 2 are on $\mathcal{A}$'s side and 3 and 4 are on $\mathcal{B}$'s side. By construction, 
$[Z^{\prime}_1, Z^{\prime}_2] = 0$, 
$[X^{\prime}_1, X^{\prime}_2] = 0$, 
$[Z^{\prime \prime}_1, X^{\prime \prime}_2] = 0$, and
$[X^{\prime \prime}_1, Z^{\prime \prime}_2] = 0$.
Also, any operators defined from $\mathcal{A}$'s measurements is on register $\mathcal{H}_A$ and analogously for $\mathcal{B}$.  Note that $V^{\prime}_3$, $W^{\prime}_3$, $V^{\prime}_3$, and $W^{\prime}_3$ do not necessarily have eigenvalues $\pm 1$, but certainly their eigenvalues are real and have absolute value at most 1.

Now suppose that we examine the CHSH correlations for each subtest and find that we achieve the quantum limit. This can be expressed as
\begin{widetext}
\begin{eqnarray}
    \frac{1}{2}\bra{\psi^{\prime}}
    \left[
    \left(
        Z^{\prime}_1 +Z^{\prime \prime}_1
    \right)    
    \left(
        V^{\prime}_3 + W^{\prime}_3
    \right)
    +
    \left(
        X^{\prime}_1 +X^{\prime \prime}_1
    \right)    
    \left(
        V^{\prime}_3 - W^{\prime}_3
    \right)
    \right]
    \ket{\psi^{\prime}}
    & = &
    2 \sqrt{2}
\\
    \frac{1}{2}\bra{\psi^{\prime}}
    \left[
    \left(
        Z^{\prime}_2 +Z^{\prime \prime}_2
    \right)    
    \left(
        V^{\prime}_4 + W^{\prime}_4
    \right)
    +
    \left(
        X^{\prime}_2 +X^{\prime \prime}_2
    \right)    
    \left(
        V^{\prime}_4 - W^{\prime}_4
    \right)
    \right]
    \ket{\psi^{\prime}}
    & = & 
    2 \sqrt{2}.
\end{eqnarray}
\end{widetext}
These can be decomposed into four separate CHSH correlations:
\begin{eqnarray}
\bra{\psi^{\prime}}
\left[
    Z^{\prime}_1
\left(
    V^{\prime}_3 + W^{\prime}_3
\right)
+
    X^{\prime}_1
\left(
    V^{\prime}_3 - W^{\prime}_3
\right)
\right]
\ket{\psi^{\prime}}
& \leq &
2 \sqrt{2} \\
\bra{\psi^{\prime}}
\left[
    Z^{\prime \prime}_1
\left(
    V^{\prime}_3 + W^{\prime}_3
\right)
+
    X^{\prime \prime}_1
\left(
    V^{\prime}_3 - W^{\prime}_3
\right)
\right]
\ket{\psi^{\prime}}
& \leq &
2 \sqrt{2}
\\
    \bra{\psi^{\prime}}
    \left[
        Z^{\prime}_2
    \left(
        V^{\prime}_4 + W^{\prime}_4
    \right)
    +
        X^{\prime}_2
    \left(
        V^{\prime}_4 - W^{\prime}_4
    \right)
    \right]
    \ket{\psi^{\prime}}
    & \leq &
    2 \sqrt{2} 
\\
    \bra{\psi^{\prime}}
    \left[
        Z^{\prime \prime}_2
    \left(
        V^{\prime}_4 + W^{\prime}_4
    \right)
    +
        X^{\prime \prime}_2
    \left(
        V^{\prime}_4 - W^{\prime}_4
    \right)
    \right]
    \ket{\psi^{\prime}}
    & \leq &
    2 \sqrt{2}. \qquad
\end{eqnarray}
In order to see the overall $2 \sqrt{2}$, equality must hold for all of the above.

We now wish to use these correlations to determine some properties of the operators.  Examining the proof of Theorem~2 in~\cite{McKague2012Robust} we can extract the following statement: 

\begin{lemma}[\cite{McKague2012Robust}]
\label{theorem:chshrobust}
Suppose that the observables $Z^{\prime}_A$, $X^{\prime}_A$, $V^{\prime}_B$ and $W^{\prime}_B$ with eigenvalues $\pm 1$, acting on a state $\ket{\psi^{\prime}} \in \mathcal{H}_A \otimes \mathcal{H}_B$, are such that
\begin{equation}
\bra{\psi^{\prime}}
\left[
    Z^{\prime}_A
\left(
    V^{\prime}_B + W^{\prime}_B
\right)
+
    X^{\prime}_A
\left(
    V^{\prime}_B - W^{\prime}_B
\right)
\right]
\ket{\psi^{\prime}}
=
2 \sqrt{2}
\end{equation}
then defining
\begin{eqnarray}
    X^{\prime}_B & = &
    \frac{
        V^{\prime}_B + W^{\prime}_B
    }{
        \left|
            V^{\prime}_B + W^{\prime}_B
        \right|
    } 
\\
    Z^{\prime}_B & = &
    \frac{
        V^{\prime}_B - W^{\prime}_B
    }{
        \left|
            V^{\prime}_B - W^{\prime}_B
        \right|
    }
\end{eqnarray}
we have
\begin{eqnarray}
Z^{\prime}_A \ket{\psi^{\prime}} & = & X^{\prime}_B \ket{\psi^{\prime}} \\
X^{\prime}_A \ket{\psi^{\prime}} & = & Z^{\prime}_B \ket{\psi^{\prime}} \\
    X^{\prime}_A
    Z^{\prime}_A 
    \ket{\psi^{\prime}}    
    & = &
    -
    Z^{\prime}_A 
    X^{\prime}_A 
    \ket{\psi^{\prime}} \\
    X^{\prime}_B
    Z^{\prime}_B 
    \ket{\psi^{\prime}}    
    & = &
    -
    Z^{\prime}_B
    X^{\prime}_B 
    \ket{\psi^{\prime}}.
\end{eqnarray}
\end{lemma}

We must be slightly careful in the above definition of $Z^{\prime}_B$ and $X^{\prime}_B$, since $|V^{\prime}_B \pm W^{\prime}_B|$ may have 0 eigenvalues.  However, to achieve the quantum limit in the CHSH tests these 0 eigenvalues must never affect the subspace that $\ket{\psi^{\prime}}$ inhabits, so we can change all 0 eigenvalues to 1 without affecting anything.  Also, the fact that $V^{\prime}_B \pm W^{\prime}_B$ and $|V^{\prime}_B \pm W^{\prime}_B|$ commute is necessary so that their quotient is well defined. With this taken care of $Z^{\prime}_B$ and $X^{\prime}_B$ have eigenvalues $\pm 1$.

Setting 
\begin{eqnarray}
    X^{\prime}_3 & = &
    \frac{
        V^{\prime}_3 + W^{\prime}_3
    }{
        \left|
            V^{\prime}_3 + W^{\prime}_3
        \right|
    } 
\\
    Z^{\prime}_3 & = &
    \frac{
        V^{\prime}_3 - W^{\prime}_3
    }{
        \left|
            V^{\prime}_3 - W^{\prime}_3
        \right|
    }
\\
    X^{\prime}_4 & = &
    \frac{
        V^{\prime}_4 + W^{\prime}_4
    }{
        \left|
            V^{\prime}_4 + W^{\prime}_4
        \right|
    } 
\\
    Z^{\prime}_4 & = &
    \frac{
        V^{\prime}_4 - W^{\prime}_4
    }{
        \left|
            V^{\prime}_4 - W^{\prime}_4
        \right|
    }    
\end{eqnarray}
Then Lemma \ref{theorem:chshrobust} gives
\begin{eqnarray}
    Z^{\prime}_1 \ket{\psi^{\prime}} & = & X^{\prime}_3 \ket{\psi^{\prime}} \\
    X^{\prime}_1 \ket{\psi^{\prime}} & = & Z^{\prime}_3 \ket{\psi^{\prime}} \\
    Z^{\prime  \prime}_1 \ket{\psi^{\prime}} & = & X^{\prime}_3 \ket{\psi^{\prime}} \\
    X^{\prime  \prime}_1 \ket{\psi^{\prime}} & = & Z^{\prime}_3 \ket{\psi^{\prime}} \\
    Z^{\prime}_2 \ket{\psi^{\prime}} & = & X^{\prime}_4 \ket{\psi^{\prime}} \\
    X^{\prime}_2 \ket{\psi^{\prime}} & = & Z^{\prime}_4 \ket{\psi^{\prime}} \\
    Z^{\prime  \prime}_2 \ket{\psi^{\prime}} & = & X^{\prime}_4 \ket{\psi^{\prime}} \\
    X^{\prime  \prime}_2 \ket{\psi^{\prime}} & = & Z^{\prime}_4 \ket{\psi^{\prime}}
\\
    X^{\prime}_1 
    Z^{\prime}_1 
    \ket{\psi^{\prime}}    
    & = & 
    -
    Z^{\prime}_1 
    X^{\prime}_1 
    \ket{\psi^{\prime}}
\\
    X^{\prime}_2 
    Z^{\prime}_2 
    \ket{\psi^{\prime}}    
    & = &
    -
    Z^{\prime}_2 
    X^{\prime}_2 
    \ket{\psi^{\prime}}
\\
    X^{\prime}_3
    Z^{\prime}_3 
    \ket{\psi^{\prime}}    
    & = & 
    -
    Z^{\prime}_3 
    X^{\prime}_3 
    \ket{\psi^{\prime}}
\\
    X^{\prime}_4 
    Z^{\prime}_4 
    \ket{\psi^{\prime}}    
    & = &
    -
    Z^{\prime}_4 
    X^{\prime}_4 
    \ket{\psi^{\prime}}.
\end{eqnarray}

Now it is quite easy to conclude that, since $[Z^{\prime \prime}_1, X^{\prime \prime}_2] = 0$ we have 
$
    Z^{\prime}_1 
    X^{\prime}_2 
    \ket{\psi^{\prime}}
=
    X^{\prime}_2 
    Z^{\prime}_1 
    \ket{\psi^{\prime}}
$ and similarly
$
    Z^{\prime}_3 
    X^{\prime}_4 
    \ket{\psi^{\prime}}
=
    X^{\prime}_3 
    Z^{\prime}_4 
    \ket{\psi^{\prime}}
$.

From the above properties we can straightforwardly prove that equation~\eqref{eq:XZswapgeneral9} holds.  For example, let $s = 1100$.  Then
\begin{eqnarray}
    X^{\prime}_1
    X^{\prime}_2
    \ket{\psi^{\prime}} 
    & = &
    X^{\prime}_1
    Z^{\prime}_4
    \ket{\psi^{\prime}} \\
    & = &
    Z^{\prime}_4
    Z^{\prime}_3
    \ket{\psi^{\prime}} \\
    & = &
    Z^{\prime}_3
    Z^{\prime}_4
    \ket{\psi^{\prime}}.
\end{eqnarray}
Other cases follow similarly.  

Now let $p,q \in \{0,1\}^4$.  We would like to conclude that 
$
    X^{\prime p} 
    Z^{\prime q} 
    \ket{\psi^{\prime}} 
    = 
    (-1)^{p \cdot q} 
    Z^{\prime q} 
    X^{\prime p} 
    \ket{\psi^{\prime}}
$.
As an example let us take $p = q = 1100$.  First, we shuffle the operators on the second qubit over to $\mathcal{B}$'s side and anticommute $X^{\prime}_1$ and $Z^{\prime}_1$:
\begin{eqnarray}
    X^{\prime}_1
    X^{\prime}_2
    Z^{\prime}_1
    Z^{\prime}_2
    \ket{\psi^{\prime}}
    & = &
    X^{\prime}_1
    X^{\prime}_2
    Z^{\prime}_1
    X^{\prime}_4
    \ket{\psi^{\prime}} \\
    & = &
    X^{\prime}_4
    X^{\prime}_1
    Z^{\prime}_1
    X^{\prime}_2
    \ket{\psi^{\prime}} \\
    & = &
    X^{\prime}_4
    Z^{\prime}_4 
    X^{\prime}_1
    Z^{\prime}_1
    \ket{\psi^{\prime}} \\
    & = &
    -
    B^{\prime}_4
    Z^{\prime}_4 
    Z^{\prime}_1
    X^{\prime}_1
    \ket{\psi^{\prime}}.
\end{eqnarray}
Next we shuffle back to $\mathcal{A}$'s side and anti-commute $X^{\prime}_2$ and $Z^{\prime}_2$:
\begin{eqnarray}
    -
    X^{\prime}_4
    Z^{\prime}_4 
    Z^{\prime}_1
    X^{\prime}_1
    \ket{\psi^{\prime}}
    & = &
    - 
    Z^{\prime}_1
    X^{\prime}_1
    X^{\prime}_2
    Z^{\prime}_2
    \ket{\psi^{\prime}} \\
    & = &
    Z^{\prime}_1
    X^{\prime}_1
    Z^{\prime}_2
    X^{\prime}_2
    \ket{\psi^{\prime}}.
\end{eqnarray}
Finally, we adjust the order
\begin{eqnarray}
    Z^{\prime}_1
    X^{\prime}_1
    Z^{\prime}_2
    X^{\prime}_2   
    \ket{\psi^{\prime}} 
    & = & 
    Z^{\prime}_4
    Z^{\prime}_1
    X^{\prime}_1
    Z^{\prime}_2
    \ket{\psi^{\prime}}  \\
    & = & 
    X^{\prime}_4
    Z^{\prime}_1
    Z^{\prime}_2
    X^{\prime}_1
    \ket{\psi^{\prime}}  \\
    & = & 
    Z^{\prime}_1
    Z^{\prime}_2
    X^{\prime}_1
    X^{\prime}_2
    \ket{\psi^{\prime}}.
\end{eqnarray}
Other cases where $p,q = ??00$ or $p,q = 00??$ proceed similarly. The general case is solved by combining the two approaches.  If $A_1 A_2 \ket{\psi^{\prime}} = a A_2 A_1 \ket{\psi^{\prime}}$ and $B_1 B_2 \ket{\psi^{\prime}} = b B_2 B_1 \ket{\psi^{\prime}}$ then we have
\begin{eqnarray}
A_1 A_2 B_1 B_2 \ket{\psi^{\prime}} & = & 
b A_1 A_2 B_2 B_1 \ket{\psi^{\prime}} \\
& = & 
b B_2 B_1 A_1 A_2 \ket{\psi^{\prime}} \\
& = & 
ab B_2 B_1 A_2 A_1 \ket{\psi^{\prime}} \\
& = & 
ab A_2 A_1 B_2 B_1  \ket{\psi^{\prime}}. 
\end{eqnarray}
Thus we have shown than equation~\eqref{eq:XZanticommute9} holds.  Now we can appeal to Lemma~\ref{lemma:graphstateselftestconditions} which tells us that there exists an isometry $\Phi = \Phi_\mathcal{A} \otimes \mathcal{B}$ and a state $\ket{junk}$ such that
\begin{equation}
\Phi \left(
    X^{\prime p} 
    Z^{\prime q} 
    \ket{\psi^{\prime}}
\right) =
\ket{junk}
X^{p}
Z^{q}
\ket{\psi}
\end{equation}
for any $p,q \in \{0,1\}^4$ meaning that we can identify two maximally entangled pairs of qubits between $\mathcal{A}$ and $\mathcal{B}$ and furthermore that the measurements used are equivalent to the ideal operators.

\section{Proof of the unitarity of the swap}
\label{swapunitary}

Here we prove that the map $$\ket{\psi}_{\mathcal{A}}\longrightarrow
\mathcal{S}_{\mathcal{A}\mathcal{A'}}\ket{\psi}_{\mathcal{A}}\otimes\ket{0,0}_{\mathcal{A'}}$$ with $\mathcal{S}_{\mathcal{A}\mathcal{A'}}$ safisfying Eq.~\eqref{SS2} preserves the scalar product. To start, we note that the action of this map can be written as
\begin{align}
\mathcal{S}_{\mathcal{A}\mathcal{A'}}\ket{\psi}_{\mathcal{A}}\otimes\ket{0,0}_{\mathcal{A'}} = \sum_{i,j=0}^1(S^{i,j}_{\mathcal{A}}\otimes \one)\ket{\psi}_{\mathcal{A}}\otimes\ket{i,j}_{\mathcal{A'}},
\end{align}
where $S^{i,j}_{\mathcal{A}}$ are operators acting on Alice's system, which are descibed in Eq.~\eqref{SS2}, i.e.
\begin{align}
S^{0,0}_\mathcal{A}&= \Pi_{0|0}\\
S^{0,1}_\mathcal{A}&= (\Pi_{0|3}-\Pi_{1|3}+\Pi_{2|3}-\Pi_{3|3})\,\Pi_{1|0}\nonumber\\
S^{1,0}_\mathcal{A}&= (\Pi_{0|3}+\Pi_{1|3}-\Pi_{2|3}-\Pi_{3|3})\,\Pi_{2|0}\nonumber\\
S^{1,1}_\mathcal{A}&= (\Pi_{0|3}-\Pi_{1|3}-\Pi_{2|3}+\Pi_{3|3})\,\Pi_{3|0}\nonumber
\end{align}

The scalar product $\braket{\phi|\psi}_\mathcal{A}$, thus becomes
\begin{align}
&\bra{\phi}_\mathcal{A}\otimes\bra{0,0}_\mathcal{A'} (S_{\mathcal{A}\mathcal{A'}})^\dagger\ S_{\mathcal{A}\mathcal{A'}}  \ket{\psi}_\mathcal{A}\otimes\ket{0,0}_\mathcal{A'}\\
&=\sum_{i,j}\bra{\phi}_\mathcal{A}\otimes\bra{i,j}_\mathcal{A'} (S^{i,j}_{\mathcal{A}}\otimes \one)^\dagger\nonumber\\
&\ \ \ \times\sum_{k,\ell}(S^{k,\ell}_{\mathcal{A}}\otimes \one)\ket{\psi}_{\mathcal{A}}\otimes\ket{k,\ell}_{\mathcal{A'}}\nonumber\\
&=\bra{\phi}_\mathcal{A}\ \sum_{i,j} (S^{i,j}_{\mathcal{A}})^\dagger\ 
S^{i,j}_{\mathcal{A}}\ \ket{\psi}_{\mathcal{A}}\nonumber\\
&=\braket{\psi|\phi}_\mathcal{A}\nonumber,
\end{align}
where in the last step we used the identity $\sum_{i,j} (S^{i,j}_{\mathcal{A}})^\dagger\ 
S^{i,j}_{\mathcal{A}}=\one$, which can be checked explicitly.

\section{Analytic bound for the Magic Square game}
\label{appmm}

In order to prove our analytic bound for the magic square game, we will appeal to \cite{MM15}, Lemma 6, stated below for the case of two singlets.  Suppose that $\mathcal{A}$ and $\mathcal{B}$ have a strategy for winning the magic square game with high probability by using some state $\ket{\psi^{\prime}}$ and some measurements.  We use this fact to derive some properties of the measurements and then appeal to \cite{MM15}, Lemma 6 in order to construct an isometry which maps $\ket{\psi^{\prime}}$ to a state close to two singlets.

\begin{lemma}[\cite{MM15}]
\label{lemma:mm15}
Suppose that 
$\ket{\psi^{\prime}} \in \mathcal{H}_A \otimes \mathcal{H}_B$ 
is a state, 
$\{X^{\prime}_1, X^{\prime}_2\}$ 
and 
$\{Z^{\prime}_1, Z^{\prime}_2\}$ 
are each a set of unitary, Hermitian, commuting operators on $\mathcal{H}_A$, and 
$\{X^{\prime}_3, X^{\prime}_4\}$ 
and
$\{Z^{\prime}_3, Z^{\prime}_4\}$ 
are each a set of unitary, Hermitian, commuting operators on $\mathcal{H}_B$, such that for all $k \neq \ell$
\begin{eqnarray}
\label{eq:XZapproxcommute2}
	\norm{
		X^{\prime}_k Z^{\prime}_\ell \ket{\psi^{\prime}}
		-
		Z^{\prime}_\ell X^{\prime}_k \ket{\psi^{\prime}}
	}
	& \leq & \epsilon_1
\\
\label{eq:XnearZ2}
	\norm{
		X^{\prime}_k \ket{\psi^{\prime}}
		-
		Z^{\prime}_{k + 2} \ket{\psi^{\prime}}
	}
	& \leq & \epsilon_2  
\\
\label{eq:XZanticommute2}
	\norm{
		Z^{\prime}_k X^{\prime}_k \ket{\psi^{\prime}}
		+ 
		X^{\prime}_k Z^{\prime}_k \ket{\psi^{\prime}}
	}
	& \leq & \epsilon_3
\end{eqnarray}
where $k+2$ is taken modulo $4$. Then there exists an isometry $\Phi = \Phi_A \otimes \Phi_B$ and a state $\ket{junk}$ such that for any $p,q \in (0,1)^{4}$
\begin{multline}
    \norm{
        \Phi(
            X^{\prime q}
            Z^{\prime p}
            \ket{\psi^{\prime}}
        )
        -
        \ket{junk}
        X^q
        Z^p
        \ket{\psi}    
    }
    \leq 
    \\
    \sqrt{
        \frac{|p|}{2}
        \left(
            3 \epsilon_1 +
            8 \epsilon_2 +
            \epsilon_3
        \right)
        + 
            \epsilon_1
        +
            \epsilon_3 
        +
            4\epsilon_2    
    } 
\\
    +
    \sqrt{
        2 \epsilon_1 +
        10 \epsilon_2 +
        2 \epsilon_3
    }
    .
\end{multline}
where $\ket{\psi}$ is two singlets between parties $\mathcal{A}$ and $\mathcal{B}$.
\end{lemma}

By way of explanation, $X = \sigma_x$ and $Z = \sigma_z$ are Pauli operators.  The operators marked with $\prime$ are unknown, save for the properties mentioned in the lemma.  Setting $p = q = 0000$ we obtain a bound on the state $\ket{\psi^{\prime}}$.  For other values of $p$ and $q$ we obtain bounds on the action of the various operators.

We can make the following correspondence with the notation used in this paper: $X^{\prime}_1 = \sigma_x^{\mathcal{A}_I}$, $X^{\prime}_2 = \sigma_x^{\mathcal{A}_{II}}$, $X^{\prime}_3 = \sigma_x^{\mathcal{B}_{I}}$, $X^{\prime}_4 = \sigma_x^{\mathcal{B}_{II}}$, and analogously for $Z$.  Further we define, for $q \in (0,1)^n$ a bit string with $q_j$ the $j$th bit of the string
\begin{equation}
X^{\prime q} :=
\prod_j
X^{\prime q_j}_j
\end{equation}
and analogously for $Z^{\prime p}$.  For bit string $p \in (0,1)^n$, $|p|$ is the number of 1s in $p$.  

The singlet is expressed in a different basis in \cite{MM15}, which is why we observe correlations between $X$ and $Z$ rather than between $X$ and $X$.  So $\ket{\psi}$ is two copies of the state $\frac{1}{2} \left(\ket{00} + \ket{01} +\ket{10} - \ket{11}\right)$.

Now we fix some strategy of $\mathcal{A}$ and $\mathcal{B}$ for playing the magic square game.  They hold onto some joint state $\ket{\psi^{\prime}}$ and perform projective measurements $\Pi^{\mathcal{A}}_{a| x}$ and $\Pi^{\mathcal{B}}_{b| z}$ for inputs $x$ and $z$ and outputs $a$ and $b$.  Next we define the operators required for Lemma~\ref{lemma:mm15}.
\begin{eqnarray}
Z^{\prime}_{1} & := & \Pi^{\mathcal{A}}_{0|0} + \Pi^{\mathcal{A}}_{1|0} - \Pi^{\mathcal{A}}_{2|0}- \Pi^{\mathcal{A}}_{3|0} \\
Z^{\prime}_{2} & := & \Pi^{\mathcal{A}}_{0|0} - \Pi^{\mathcal{A}}_{1|0} + \Pi^{\mathcal{A}}_{2|0} - \Pi^{\mathcal{A}}_{3|0} \\
X^{\prime}_1 & := & \Pi^{\mathcal{A}}_{0|1} + \Pi^{\mathcal{A}}_{1|1} - \Pi^{\mathcal{A}}_{2|1}- \Pi^{\mathcal{A}}_{3|1} \\
X^{\prime}_2 & := & \Pi^{\mathcal{A}}_{0|1} - \Pi^{\mathcal{A}}_{1|1} + \Pi^{\mathcal{A}}_{2|1} - \Pi^{\mathcal{A}}_{3|1} \\
Z^{\prime}_{3} & := & \Pi^{\mathcal{B}}_{0|0} + \Pi^{\mathcal{B}}_{1|0} - \Pi^{\mathcal{B}}_{2|0}- \Pi^{\mathcal{B}}_{3|0} \\
Z^{\prime}_{4} & := & \Pi^{\mathcal{B}}_{0|1} - \Pi^{\mathcal{B}}_{1|1} + \Pi^{\mathcal{B}}_{2|1} - \Pi^{\mathcal{B}}_{3|1} \\
X^{\prime}_3 & := & \Pi^{\mathcal{B}}_{0|1} - \Pi^{\mathcal{B}}_{1|1} + \Pi^{\mathcal{B}}_{2|1}- \Pi^{\mathcal{B}}_{3|1} \\
X^{\prime}_4 & := & \Pi^{\mathcal{B}}_{0|0} - \Pi^{\mathcal{B}}_{1|0} + \Pi^{\mathcal{B}}_{2|0} - \Pi^{\mathcal{B}}_{3|0} \\
W^{\prime}_1 & := & \Pi^{\mathcal{A}}_{0|2} + \Pi^{\mathcal{A}}_{1|2} - \Pi^{\mathcal{A}}_{2|2} - \Pi^{\mathcal{A}}_{3|2} \\
W^{\prime}_2 & := & \Pi^{\mathcal{A}}_{0|2} - \Pi^{\mathcal{A}}_{1|2} + \Pi^{\mathcal{A}}_{2|2} - \Pi^{\mathcal{A}}_{3|2} \\
W^{\prime}_3 & := & \Pi^{\mathcal{B}}_{0|2} + \Pi^{\mathcal{B}}_{1|2} - \Pi^{\mathcal{B}}_{2|2} - \Pi^{\mathcal{B}}_{3|2} \\
W^{\prime}_4 & := & \Pi^{\mathcal{B}}_{0|2} - \Pi^{\mathcal{B}}_{1|2} + \Pi^{\mathcal{B}}_{2|2} - \Pi^{\mathcal{B}}_{3|2} 
\end{eqnarray}
In all cases sub-indices 1 and 2 are on $\mathcal{A}$'s side and 3 and 4 are on $\mathcal{B}$'s side. By construction, $[Z^{\prime}_1, Z^{\prime}_2] = 0$, $[X^{\prime}_1, X^{\prime}_2] = 0$, 
$[Z^{\prime}_3, X^{\prime}_4] = 0$,
$[X^{\prime}_3, Z^{\prime}_4] = 0$,
$[W^{\prime}_1, W^{\prime}_2] = 0$ and
$[W^{\prime}_3, W^{\prime}_4] = 0$.  Also, any operators defined from $\mathcal{A}$'s measurements is on register $\mathcal{H}_A$ and analogously for $\mathcal{B}$.  These are the commutation relationships are required by Lemma~\ref{lemma:mm15}.

We can express the condition that this is a perfect strategy in terms of expectation values of the above defined operators.  For example, 
$\bra{\psi^{\prime}} 
    Z^{\prime}_1 
    X^{\prime}_3
\ket{\psi^{\prime}}
= 1
$.  To obtain a robust bound, let us suppose that this strategy is uniformly close to perfect, meaning that the expectation values are within $\epsilon$ of perfect on all conditions.  This gives us the following list of conditions:
\begin{eqnarray}
\label{eq:zi}
\bra{\psi^{\prime}}
    Z^{\prime}_1 
    X^{\prime}_3
\ket{\psi^{\prime}}
& \geq & 1 - \epsilon \\
\label{eq:zii}
\bra{\psi^{\prime}}
    Z^{\prime}_2 
    X^{\prime}_4
\ket{\psi^{\prime}}
& \geq & 1 - \epsilon \\
\label{eq:xi}
\bra{\psi^{\prime}}
    X^{\prime}_1 
    Z^{\prime}_3
\ket{\psi^{\prime}}
& \geq & 1 - \epsilon \\
\label{eq:xii}
\bra{\psi^{\prime}}
    X^{\prime}_2 
    Z^{\prime}_4
\ket{\psi^{\prime}}
& \geq & 1 - \epsilon \\
\label{eq:t2b}
\bra{\psi^{\prime}}
    Z^{\prime}_1
    Z^{\prime}_2 
    W^{\prime}_3 
\ket{\psi^{\prime}}
& \geq & 1 - \epsilon \\
\label{eq:t1b}
\bra{\psi^{\prime}}
    X^{\prime}_1
    X^{\prime}_2 
    W^{\prime}_4 
\ket{\psi^{\prime}}
& \geq & 1 - \epsilon \\
\label{eq:t1a}
\bra{\psi^{\prime}}
    W^{\prime}_1 
    X^{\prime}_3 
    Z^{\prime}_4
\ket{\psi^{\prime}}
& \geq & 1 - \epsilon \\
\label{eq:t2a}
\bra{\psi^{\prime}}
    W^{\prime}_2 
    X^{\prime}_4
    Z^{\prime}_3     
\ket{\psi^{\prime}}
& \geq & 1 - \epsilon \\
\label{eq:tt}
-
\bra{\psi^{\prime}}
    W^{\prime}_1 
    W^{\prime}_2 
    W^{\prime}_3 
    W^{\prime}_4 
\ket{\psi^{\prime}}
& \geq & 1 - \epsilon\,.
\end{eqnarray}
These conditions can easily be converted into norms -- for example 
$\norm{
    Z^{\prime}_1\ket{\psi^{\prime}} 
    -
    X^{\prime}_3\ket{\psi^{\prime}}} 
\geq \sqrt{2 \epsilon}
$ -- 
by substituting the above inner products into the definition of $\norm{\cdot}$.  Then~\eqref{eq:zi} through to~\eqref{eq:xii} allow us to set $\epsilon_2 = \sqrt{2 \epsilon}$ for Lemma~\ref{lemma:mm15}.

Now let us prove that $Z^{\prime}_1$ approximately anti-commutes with $X^{\prime}_1$ to find $\epsilon_3$.  We start with \eqref{eq:tt}, which gives us
\begin{equation}
\norm{
    W^{\prime}_1 
    W^{\prime}_2 
    \ket{\psi^{\prime}}
    +
    W^{\prime}_3 
    W^{\prime}_4 
    \ket{\psi^{\prime}}
} \leq \sqrt{2 \epsilon}.
\end{equation}
Using \eqref{eq:t2a} and \eqref{eq:t1b} to estimate $W^{\prime}_2 \ket{\psi^{\prime}}$ and $W^{\prime}_4 \ket{\psi^{\prime}}$, plus the triangle inequality, we obtain
\begin{equation}
\norm{
    W^{\prime}_1 
    Z^{\prime}_4
    X^{\prime}_3    
    \ket{\psi^{\prime}}
    +
    W^{\prime}_3 
    X^{\prime}_2
    X^{\prime}_1    
    \ket{\psi^{\prime}}
} \leq 3\sqrt{2 \epsilon}.
\end{equation}
Now $W^{\prime}_1$ can be commuted to the right since it is on $\mathcal{A}$'s system while $Z^{\prime}_4X^{\prime}_3$ is on $\mathcal{B}$'s. Similarly, $W^{\prime}_3$ moves to the right.  Then we use \eqref{eq:t1a} and \eqref{eq:t2b} to estimate $W^{\prime}_1 \ket{\psi^{\prime}}$ and $W^{\prime}_3 \ket{\psi^{\prime}}$ and obtain
\begin{equation}
\norm{
    X^{\prime}_4 
    Z^{\prime}_3
    X^{\prime}_3    
    Z^{\prime}_4
    \ket{\psi^{\prime}}
    +
    X^{\prime}_2
    X^{\prime}_1
    Z^{\prime}_1
    Z^{\prime}_2
    \ket{\psi^{\prime}}
} \leq 5\sqrt{2 \epsilon}.
\end{equation}
Now, using \eqref{eq:zii} and \eqref{eq:xii} we change $Z^{\prime}_4$ to $X^{\prime}_2$ and $Z^{\prime}_2$ to $X^{\prime}_4$:
\begin{equation}
\norm{
    X^{\prime}_4      
    Z^{\prime}_3
    X^{\prime}_3     
    X^{\prime}_2
    \ket{\psi^{\prime}}
    +
    X^{\prime}_2
    X^{\prime}_1
    Z^{\prime}_1
    X^{\prime}_4
    \ket{\psi^{\prime}}
} \leq 7\sqrt{2 \epsilon}.
\end{equation}
We then multiply on the left by $X^{\prime}_2$ and $X^{\prime}_4$ to get
\begin{equation}
\norm{
    Z^{\prime}_3    
    X^{\prime}_3
    \ket{\psi^{\prime}}
    +
    X^{\prime}_1
    Z^{\prime}_1
    \ket{\psi^{\prime}}
} \leq 7\sqrt{2 \epsilon}.
\end{equation}
Finally, using \eqref{eq:zi}, commuting the resulting $Z^{\prime}_1$ to the left, and then using \eqref{eq:xi} we obtain
\begin{equation}
\norm{
    Z^{\prime}_1    
    X^{\prime}_1
    \ket{\psi^{\prime}}
    +
    X^{\prime}_1
    Z^{\prime}_1
    \ket{\psi^{\prime}}
} \leq 9\sqrt{2 \epsilon}.
\end{equation}
We can also obtain the analogous results for any $X^{\prime}_j$ and $Z^{\prime}_j$ by suitably modifying the argument.  This sets $\epsilon_3 = 9 \sqrt{2 \epsilon}$ for Lemma~\ref{lemma:mm15}.

Now we turn our attention to $Z^{\prime}_1$ and $X^{\prime}_2$ to find $\epsilon_1$.  We start with \eqref{eq:zi} and \eqref{eq:xii}, finding
\begin{equation}
\norm{
    Z^{\prime}_1
    X^{\prime}_2    
    \ket{\psi^{\prime}} 
    -
    X^{\prime}_4    
    Z^{\prime}_3
    \ket{\psi^{\prime}}
} \leq 2\sqrt{2 \epsilon}.
\end{equation}
Using 
$
[
    X^{\prime}_4
    ,
    Z^{\prime}_3
] = 0
$
and \eqref{eq:zi} and \eqref{eq:xii} again, we obtain
\begin{equation}
\norm{
    Z^{\prime}_1
    X^{\prime}_2    
    \ket{\psi^{\prime}} 
    -
    X^{\prime}_2    
    Z^{\prime}_1
    \ket{\psi^{\prime}}
} \leq 4\sqrt{2 \epsilon}.
\end{equation}
Analogously, we obtain similar relationships for the pairs 
$(X^{\prime}_1, Z^{\prime}_2)$,
$(X^{\prime}_3, X^{\prime}_4)$ and
$(Z^{\prime}_3, Z^{\prime}_4)$, setting $\epsilon_1 = 4 \sqrt{2 \epsilon}$ for Lemma~\ref{lemma:mm15}.

Now we can appeal to Lemma~\ref{lemma:mm15} to find that there exists an isometry $\Phi$ and junk state $\ket{junk}$ such that for any $p,q \in (0,1)^4$
\begin{multline}
\norm{
    \Phi(X^{\prime q}Z^{\prime p} \ket{\psi^{\prime}})
    -
    \ket{junk}
    X^q Z^p 
    \ket{\psi}
} 
\\
\leq
\left(
\sqrt{\frac{29 |p|}{2} + 17}
+
6
\right)
(2\epsilon)^{\frac{1}{4}}
\end{multline}
where $\ket{\psi}$ is two maximally entangled pairs.  This concludes our derivation of the analytic bound for the magic square game.

\end{document}


\section{Proof of the unitarity of the swap}
\label{swapunitary}

First, we write down the 1-qubit swap,
\begin{align}
\label{1swap}
	\mathcal{S}^1=&\ket{0}\bra{0}\otimes \pi_{z=0}+\ket{0}\bra{1}\otimes \pi_{z=1}\sigma_x\nonumber\\
	+&\ket{1}\bra{0}\otimes \sigma_x \pi_{z=1}+\ket{1}\bra{1}\otimes \sigma_x \pi_{z=0} \sigma_x
\end{align}

In general, we need to prove that the swap we choose for the global system is unitary. Without loosing generality, we only show the proof for one party 2-qubit swap, since the global swap is just the tensor product of two 2-qubit swap. Hence, we need to prove that for 2-qubit swap 
\begin{align}
	\mathcal{S}^\dagger\mathcal{S}=\one.
\end{align}
Since we are considering 2-qubit swap here, the swap matrix $\mathcal{S}$, which is the 2 times tensor product of (\ref{1swap}), should be a $ 4 \times 4 $ matrix.

To exam the unitarity of the matrix, what we can do is to check each element of the matrix $\mathcal{S}^\dagger\mathcal{S}$.

Let us say $(\mathcal{S}^\dagger\mathcal{S})_{11}$,
\begin{align}
	&(\mathcal{S}^\dagger\mathcal{S})_{11}\nonumber\\
	=&\mathcal{S}_{11}^\dagger\mathcal{S}_{11}+\mathcal{S}_{12}^\dagger\mathcal{S}_{21}+\mathcal{S}_{13}^\dagger\mathcal{S}_{31}+\mathcal{S}_{14}^\dagger\mathcal{S}_{41}\nonumber\nonumber\\
	=&(\mathcal{S}_{11})^\dagger\mathcal{S}_{11}+(\mathcal{S}_{21})^\dagger\mathcal{S}_{21}+(\mathcal{S}_{31})^\dagger\mathcal{S}_{31}+(\mathcal{S}_{41})^\dagger\mathcal{S}_{41}\nonumber,
\end{align}
since $\mathcal{S}_{11}$ corresponds to element $\ket{00}\bra{00}$, $\mathcal{S}_{21}$ to $\ket{01}\bra{00}$, $\mathcal{S}_{21}$ to $\ket{10}\bra{00}$ and $\mathcal{S}_{41}$ to $\ket{11}\bra{00}$, we substitute $\mathcal{S}_{i1}$ defined with (\ref{def}) and get
\begin{align}
	&(\mathcal{S}^\dagger\mathcal{S})_{11}\nonumber\\
	=&(\Pi^1_{0})^2\nonumber\\
	+&\Pi^1_{1}(\Pi^4_0-\Pi^4_1+\Pi^4_2+\Pi^4_3)(\Pi^4_0-\Pi^4_1+\Pi^4_2+\Pi^4_3)\Pi^1_{1}\nonumber\\
	+&\Pi^1_{2}(\Pi^4_0+\Pi^4_1-\Pi^4_2-\Pi^4_3)(\Pi^4_0+\Pi^4_1-\Pi^4_2-\Pi^4_3)\Pi^1_{2}\nonumber\\
	+&\Pi^1_{3}(\Pi^4_0-\Pi^4_1+\Pi^4_2+\Pi^4_3)(\Pi^4_0+\Pi^4_1-\Pi^4_2-\Pi^4_3)\nonumber\\
	&(\Pi^4_0+\Pi^4_1-\Pi^4_2-\Pi^4_3)(\Pi^4_0-\Pi^4_1+\Pi^4_2+\Pi^4_3)\Pi^1_{3}\nonumber\\
	=&\Pi^1_{0}+\Pi^1_{1}+\Pi^1_{2}+\Pi^1_{3}\nonumber\\
	=& 1.
\end{align}
Following the same way, we can also prove that the other elements of $\mathcal{S}^{\dagger}\mathcal{S}$ are exaltedly the same as an identity matrix.

\section{Analytic bound for the Magic Square game}
The strategy for this section is as follows:  (1) develop a model for strategies that are `close to' perfect, (2) derive commuting and anti-commuting properties of operators in the model (3) appeal to a lemma of McKague (\cite{MM15}, Lemma 6) which provides a suitable isometry and error bound.  Because \cite{MM15} works in terms of graph states, we will show that Alice and Bob's state is close to two copies of the isolated edge state $\frac{1}{2}\left(\ket{00} + \ket{01} + \ket{10} - \ket{11}\right)$, which is equivalent to a singlet.

Let us begin by modelling Alice and Bob's strategy.  They hold onto some joint state $\ket{\psi^{\prime}}$ and perform projective measurements $\Pi^{\mathcal{A}}_{a| x}$ (Alice) and $\Pi^{\mathcal{A}}_{b| z}$ (Bob).  For our convenience we will define some new operators:
\begin{eqnarray}
Z^{\prime}_{1} & = & \Pi^{\mathcal{A}}_{0|0} + \Pi^{\mathcal{A}}_{1|0} - \Pi^{\mathcal{A}}_{2|0}- \Pi^{\mathcal{A}}_{3|0} \\
Z^{\prime}_{2} & = & \Pi^{\mathcal{A}}_{0|0} - \Pi^{\mathcal{A}}_{1|0} + \Pi^{\mathcal{A}}_{2|0} - \Pi^{\mathcal{A}}_{3|0} \\
X^{\prime}_1 & = & \Pi^{\mathcal{A}}_{0|1} + \Pi^{\mathcal{A}}_{1|1} - \Pi^{\mathcal{A}}_{2|1}- \Pi^{\mathcal{A}}_{3|1} \\
X^{\prime}_2 & = & \Pi^{\mathcal{A}}_{0|1} - \Pi^{\mathcal{A}}_{1|1} + \Pi^{\mathcal{A}}_{2|1} - \Pi^{\mathcal{A}}_{3|1} \\
Z^{\prime}_{3} & = & \Pi^{\mathcal{B}}_{0|0} + \Pi^{\mathcal{B}}_{1|0} - \Pi^{\mathcal{B}}_{2|0}- \Pi^{\mathcal{B}}_{3|0} \\
Z^{\prime}_{4} & = & \Pi^{\mathcal{B}}_{0|1} - \Pi^{\mathcal{B}}_{1|1} + \Pi^{\mathcal{B}}_{2|1} - \Pi^{\mathcal{B}}_{3|1} \\
X^{\prime}_3 & = & \Pi^{\mathcal{B}}_{0|1} - \Pi^{\mathcal{B}}_{1|1} + \Pi^{\mathcal{B}}_{2|1}- \Pi^{\mathcal{B}}_{3|1} \\
X^{\prime}_4 & = & \Pi^{\mathcal{B}}_{0|0} - \Pi^{\mathcal{B}}_{1|0} + \Pi^{\mathcal{B}}_{2|0} - \Pi^{\mathcal{B}}_{3|0} \\
W^{\prime}_1 & = & \Pi^{\mathcal{A}}_{0|3} + \Pi^{\mathcal{A}}_{1|3} - \Pi^{\mathcal{A}}_{2|3} - \Pi^{\mathcal{A}}_{3|3} \\
W^{\prime}_2 & = & \Pi^{\mathcal{A}}_{0|3} - \Pi^{\mathcal{A}}_{1|3} + \Pi^{\mathcal{A}}_{2|3} - \Pi^{\mathcal{A}}_{3|3} \\
W^{\prime}_3 & = & \Pi^{\mathcal{B}}_{0|3} + \Pi^{\mathcal{B}}_{1|3} - \Pi^{\mathcal{B}}_{2|3} - \Pi^{\mathcal{B}}_{3|3} \\
W^{\prime}_4 & = & \Pi^{\mathcal{B}}_{0|3} - \Pi^{\mathcal{B}}_{1|3} + \Pi^{\mathcal{B}}_{2|3} - \Pi^{\mathcal{B}}_{3|3} 
\end{eqnarray}
Here the $\prime$ serves to remind us that these are not the Pauli $X$ and $Z$ operators, although we will find that they play an analogous role.  In all cases sub-indices 1 and 2 are on Alice's side and 3 and 4 are on Bob's side. By construction, $[Z^{\prime}_1, Z^{\prime}_2] = 0$, $[X^{\prime}_1, X^{\prime}_2] = 0$, 
$[Z^{\prime}_3, X^{\prime}_4] = 0$,
$[X^{\prime}_3, Z^{\prime}_4] = 0$,
$[W^{\prime}_1, W^{\prime}_2] = 0$ and
$[W^{\prime}_3, W^{\prime}_4] = 0$.  Also, any pair of operators where one is on Alice's side and the other is on Bob's side also commute.  These commuting relationships are required by \cite{MM15}, Lemma 6.

We can express the condition that this is a perfect strategy in terms of expectation values of the above defined operators.  For example, 
$\bra{\psi^{\prime}} 
    Z^{\prime}_1 
    X^{\prime}_3
\ket{\psi^{\prime}}
= 1
$.  To obtain a robust bound, let us suppose that this strategy is uniformly close to perfect, meaning that the expectation values are within $\epsilon$ of perfect on all conditions.  This gives us the following list of conditions:
\begin{eqnarray}
\label{eq:zi}
\bra{\psi^{\prime}}
    Z^{\prime}_1 
    X^{\prime}_3
\ket{\psi^{\prime}}
& \geq & 1 - \epsilon \\
\label{eq:zii}
\bra{\psi^{\prime}}
    Z^{\prime}_2 
    X^{\prime}_4
\ket{\psi^{\prime}}
& \geq & 1 - \epsilon \\
\label{eq:xi}
\bra{\psi^{\prime}}
    X^{\prime}_1 
    Z^{\prime}_3
\ket{\psi^{\prime}}
& \geq & 1 - \epsilon \\
\label{eq:xii}
\bra{\psi^{\prime}}
    X^{\prime}_2 
    Z^{\prime}_4
\ket{\psi^{\prime}}
& \geq & 1 - \epsilon \\
\label{eq:t2b}
\bra{\psi^{\prime}}
    Z^{\prime}_1
    Z^{\prime}_2 
    W^{\prime}_3 
\ket{\psi^{\prime}}
& \geq & 1 - \epsilon \\
\label{eq:t1b}
\bra{\psi^{\prime}}
    X^{\prime}_1
    X^{\prime}_2 
    W^{\prime}_4 
\ket{\psi^{\prime}}
& \geq & 1 - \epsilon \\
\label{eq:t1a}
\bra{\psi^{\prime}}
    W^{\prime}_1 
    X^{\prime}_3 
    Z^{\prime}_4
\ket{\psi^{\prime}}
& \geq & 1 - \epsilon \\
\label{eq:t2a}
\bra{\psi^{\prime}}
    W^{\prime}_2 
    X^{\prime}_4
    Z^{\prime}_3     
\ket{\psi^{\prime}}
& \geq & 1 - \epsilon \\
\label{eq:tt}
-
\bra{\psi^{\prime}}
    W^{\prime}_1 
    W^{\prime}_2 
    W^{\prime}_3 
    W^{\prime}_4 
\ket{\psi^{\prime}}
& \geq & 1 - \epsilon
\end{eqnarray}
These conditions can easily be converted into norms -- for example 
$\norm{
    Z^{\prime}_1\ket{\psi^{\prime}} 
    -
    X^{\prime}_3\ket{\psi^{\prime}}} 
\geq \sqrt{2 \epsilon}
$ -- 
by substituting the above inner products into the definition of $\norm{\cdot}$.  Then~\eqref{eq:zi} through to~\eqref{eq:xii} allow us to set $\epsilon_2 = \sqrt{2 \epsilon}$ for \cite{MM15}, Lemma 6.

Now let us prove that $Z^{\prime}_1$ approximately anti-commutes with $X^{\prime}_1$.  We start with \eqref{eq:tt}, which gives us
\begin{equation}
\norm{
    W^{\prime}_1 
    W^{\prime}_2 
    \ket{\psi^{\prime}}
    +
    W^{\prime}_3 
    W^{\prime}_4 
    \ket{\psi^{\prime}}
} \leq \sqrt{2 \epsilon}.
\end{equation}
Using \eqref{eq:t2a} and \eqref{eq:t1b} plus the triangle inequality we obtain
\begin{equation}
\norm{
    W^{\prime}_1 
    Z^{\prime}_4
    X^{\prime}_3    
    \ket{\psi^{\prime}}
    +
    W^{\prime}_3 
    X^{\prime}_2
    X^{\prime}_1    
    \ket{\psi^{\prime}}
} \leq 3\sqrt{2 \epsilon}.
\end{equation}
We can do a similar thing again with \eqref{eq:t1a} and \eqref{eq:t2b}, first noting that $W^{\prime}_1$ can first be commuted to the right, as can $    W^{\prime}_3$.  We obtain
\begin{equation}
\norm{
    X^{\prime}_4 
    Z^{\prime}_3
    X^{\prime}_3    
    Z^{\prime}_4
    \ket{\psi^{\prime}}
    +
    X^{\prime}_2
    X^{\prime}_1
    Z^{\prime}_1
    Z^{\prime}_2
    \ket{\psi^{\prime}}
} \leq 3\sqrt{2 \epsilon}.
\end{equation}
Now, using \eqref{eq:zii} and \eqref{eq:xii} we find
\begin{equation}
\norm{
    X^{\prime}_4      
    Z^{\prime}_3
    X^{\prime}_3     
    X^{\prime}_2
    \ket{\psi^{\prime}}
    +
    X^{\prime}_2
    X^{\prime}_1
    Z^{\prime}_1
    X^{\prime}_4
    \ket{\psi^{\prime}}
} \leq 5\sqrt{2 \epsilon}
\end{equation}
and then multiply everything by $X^{\prime}_2$ and $X^{\prime}_4$ to get
\begin{equation}
\norm{
    Z^{\prime}_3    
    X^{\prime}_3
    \ket{\psi^{\prime}}
    +
    X^{\prime}_1
    Z^{\prime}_1
    \ket{\psi^{\prime}}
} \leq 5\sqrt{2 \epsilon}.
\end{equation}
Finally, using \eqref{eq:zi}, commuting the resulting $Z^{\prime}_1$ to the left, and then using \eqref{eq:xi} we obtain
\begin{equation}
\norm{
    Z^{\prime}_1    
    X^{\prime}_1
    \ket{\psi^{\prime}}
    +
    X^{\prime}_1
    Z^{\prime}_1
    \ket{\psi^{\prime}}
} \leq 7\sqrt{2 \epsilon}.
\end{equation}
We can also obtain the analogous results for any $X^{\prime}_j$ and $Z^{\prime}_j$ by suitably modifying the argument.  This sets $\epsilon_3 = 7 \sqrt{2 \epsilon}$ for \cite{MM15}, Lemma 6.

Now we turn our attention to $Z^{\prime}_1$ and $X^{\prime}_2$.  So far we do not know how these relate.  We start with \eqref{eq:zi} and \eqref{eq:xii}, finding
\begin{equation}
\norm{
    Z^{\prime}_1
    X^{\prime}_2    
    \ket{\psi^{\prime}} 
    -
    X^{\prime}_4    
    Z^{\prime}_3
    \ket{\psi^{\prime}}
} \leq 2\sqrt{2 \epsilon}.
\end{equation}
Using 
$
[
    X^{\prime}_4
    ,
    Z^{\prime}_3
] = 0
$
and \eqref{eq:zi} and \eqref{eq:xii} again, we obtain
\begin{equation}
\norm{
    Z^{\prime}_1
    X^{\prime}_2    
    \ket{\psi^{\prime}} 
    -
    X^{\prime}_2    
    Z^{\prime}_1
    \ket{\psi^{\prime}}
} \leq 4\sqrt{2 \epsilon}.
\end{equation}
Analogously, we obtain similar relationships for the pairs 
$(X^{\prime}_1, Z^{\prime}_2)$,
$(X^{\prime}_3, X^{\prime}_4)$ and
$(Z^{\prime}_3, Z^{\prime}_4)$, setting $\epsilon_1 = 4 \sqrt{2 \epsilon}$ in \cite{MM15}, Lemma 6.

Now we can appeal to \cite{MM15}, Lemma 6 to find that there exists an isometry $\Phi$ and junk state $\ket{junk}$ such that
\begin{eqnarray}
\nonumber
\norm{
    \Phi(X^{\prime q}Z^{\prime p} \ket{\psi^{\prime}})
    -
    \ket{junk}
    X^q Z^p 
    \ket{\psi}
} 
\qquad \qquad
\\
\qquad \qquad
\qquad \qquad
\leq
\left(
\sqrt{\frac{21 |p|}{2}}
+
2
\sqrt{7}
\right)
(2\epsilon)^{\frac{1}{4}}
\end{eqnarray}
where $\ket{\psi}$ is the graph state with two isolated edges, i.e.\ two maximally entangled pairs.  Here $p$ and $q$ are two 4-bit strings, $|p|$ is the number of 1s in $p$, and
\begin{equation}
X^{\prime q} =
\prod_j
X^{\prime q_j}_j
\end{equation}
(and analogously for $Z^{\prime p}$).